\documentclass{article}
\usepackage{frascatiphys}
\usepackage{graphicx}
\begin{document}
\title{SPECTRAL FEATURES IN GALACTIC COSMIC RAYS}

%

\author{
Manuela Vecchi\\
{\em KVI - Center for Advanced Radiation Technology, University of Groningen,}\\ 
{\em The Netherlands and S\~ao Carlos Institute of Physics, University of S\~ao Paulo, Brasil} \\
}%

\maketitle
\baselineskip=11.6pt
\begin{abstract}
Recent results by space borne experiments took cosmic ray data to a precision level. These new  results are able to challenge 
the conventional scenario for cosmic ray acceleration and propagation in the Milky Way.
In these contributions, written for the XVII Vulcano Workshop, we will give an overview of the latest results of the cosmic ray fluxes, and some possible interpretations 
will be discussed. These measurements have a common feature, namely the presence of 
unexpected and still not yet fully understood spectral features.		
\end{abstract}
\baselineskip=14pt

\section{Introduction}
We are in a very exciting phase for the field of astroparticle physics: the observation of gravitational waves from the merger 
of a binary neutron star system~\cite{NS} in coincidence with the 
electromagnetic radiation detected in a broad range of wavelenghts, in August 2017, 
marked a milestone for multi-messenger astronomy~\cite{NSmm}, while the IceCube Collaboration announced, in 
July 2018, the first evidence for a source of high-energy (TeV) cosmic neutrinos~\cite{IC}. 
The  BL Lac object TXS 0506+056 is likely to be the first identified source of high energy neutrinos and, consequently, of cosmic rays~\cite{nucr}. 

More than 100 years after the discovery of V. Hess, the understanding of the origin, the acceleration and 
propagation mechanisms of cosmic rays (CRs) in the galaxy and beyond is not yet completely understood and 
cosmic ray physics is still a lively and fascinating field of research. 

In the simplified ``conventional scenario''~\cite{rev} to describe the origin and propagation of CRs up to the knee, the 
primary CRs (e.g. H, He, C) are accelerated in 
Supernova remnants (SNR) via diffusive shock acceleration up to PeV energies, while their propagation
in the interstellar medium (ISM)  is described by an
homogeneous and energy-dependent diffusion coefficient $K$.  
Once primary CRs are released from the sources, they propagate in the 
 interstellar medium (ISM), made mainly by protons and helium nuclei, where they are 
 confined by the magnetic fields for times of the order of  a few million years~\cite{blasiRev}.
 When primary particles interact with the ISM they produce 
 secondary CRs, like lithium, beryllium, boron as well as antimatter particles such as positrons and antiprotons.
This theoretical framework provides featureless and universal (species independent) single power-law energy spectra, and
it 
was supported by experimental results up to one decade ago.

This work aims at providing a concise description of the latest experimental results on direct CR measurements up to the knee, 
and to provide a quick overview of possible interpretation scenarios. 
As a further reading, we suggest the reviews by P.~Serpico~\cite{Serpico2015} and L.~Drury~\cite{Drury2017}.

\section{Spectral features in galactic cosmic ray measurements}
The study of spectral features in the fluxes of 
galactic CRs will provide us with a deeper understanding of the physical processes that occur in the Milky Way. 

\begin{figure}[h!]
    \begin{center}
        {\includegraphics[scale=0.5]{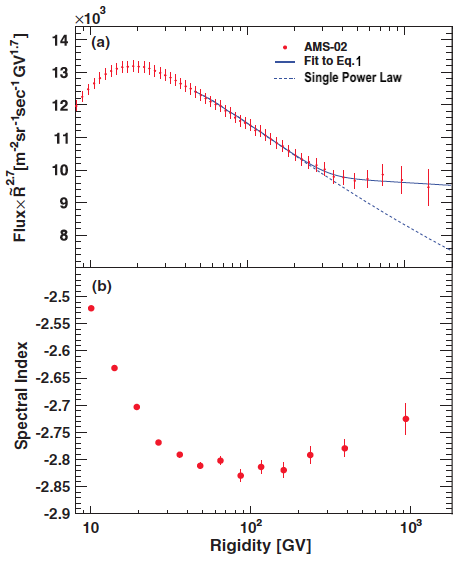}}
        \caption{ \emph{Top plot:} The AMS proton flux multiplied by $R^{2.7}$ as a
function of rigidity. The solid curve indicates the fit of equation 1
to the data. The dashed curve indicates the single power law behavior. \emph{Bottom plot:}
The flux spectral index $\gamma$ as a function of rigidity. Plots adapted from~\cite{proton}. }
\label{exp:fig1}
    \end{center}
\end{figure}

\begin{figure}[htb]
    \begin{center}
        {\includegraphics[scale=0.4]{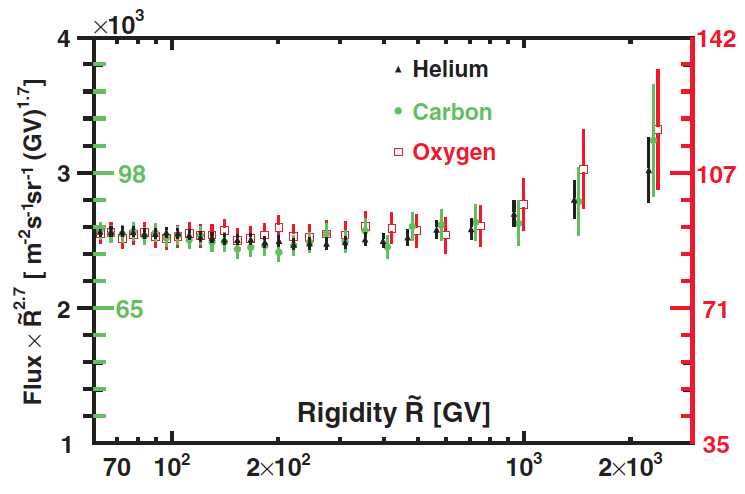}}
        \caption{The rigidity dependence of the helium (left black axis),
carbon (left green axis), and oxygen (right red axis) fluxes~\cite{heco}. For
clarity, horizontal positions of the helium and oxygen data points
above 400 GV are displaced with respect to the carbon.}
\label{exp:fig2}
    \end{center}
\end{figure}

The ``conventional model'' was a reasonable option to describe the CR data until the beginning of 2000, when detectors with large acceptance and good resolution 
were brought to the uppermost layers of the atmosphere or to space: the first hints of deviations from the 
single power law were provided by the CREAM balloon experiment~\cite{cream}, which suggested an indication for a transition in the spectral index of CR 
proton, helium and heavier nuclei. However, the large uncertainties prevented for an unambigous claim. The 
PAMELA Collaboration published in 2011
precise measurement of proton and helium fluxes~\cite{PAMELApHe} between 1 GV to 1.2 TV, showing a clear feature above 200 GeV. 
The AMS-02 collaboration in 2015~\cite{proton}~\cite{heliumF} showed  
that both the proton and helium spectra
 cannot be described as a single power law (between 1 GV to 1.8 TV), and that
a transition in the spectral index takes place above 200 GeV. 
The top panel of figure~\ref{exp:fig1} shows the flux of CR protons measured by AMS-02 (red dots) as a function of 
rigidity~\footnote{The rigidity is given by the particle momentum over the charge $R=\frac{pc}{Ze}$}, between 1 GV and 2 TeV. The single power law behaviour, namely
$\Phi(R)~= C R^{-\gamma}$, is displayed in 
the dashed line. 
The transition in the spectral index occurs above 200 GV, and can be 
described using 5 parameters, as follows: 
\begin{equation}
\Phi(R)= C R^{-\gamma} \Big( 1 + \frac{R}{R_0}^{\Delta \gamma/s} \Big)^s
\end{equation}
where the parameter $s$ quantifies  the smoothness of the transition of the
spectral index, from $\gamma$ to $\gamma+\Delta \gamma$, that occurs at the rigidity $R_0$. The fit obtained in~\cite{proton} yelds $\gamma \sim -2.814$
for a transition rigidity value $R_0 \sim 366 ~GV$ and $\Delta \gamma \sim 0.133$.
The result of the fit to this function is shown in the solid line in the top panel of figure~\ref{exp:fig1}.
The spectral index as a function of rigidity is
shown in the bottom panel of figure~\ref{exp:fig1}.
The helium flux was also found to show puzzling spectral features. Not only the helium flux cannot be described by a single power law, but the ratio
between the proton and the helium flux is rigidity-dependent. This behavior is not expected in the conventional model, and it was also observed in the flux ratio of other species,
like C/p, O/p~\cite{heco}. This can be inferred from figure~\ref{exp:fig2} that shows that helium, carbon and oxygen exibit the same rigidity 
dependence, with different abundances. The black dots in figure~\ref{exp:fig2} show the CR helium flux as a function of rigidity, while the green and red dots show the flux of heavier CR species: carbon
and oxygen. The three 
species are mainly primary particles. 

The AMS-02 recently published precise measurements of lithium, berillium and boron~\cite{libeb}, mostly secondary CR species, reporting that 
the three fluxes deviate from a single power law above 200 GV in an identical way. 
\begin{figure}[htb]
    \begin{center}
        {\includegraphics[scale=0.5]{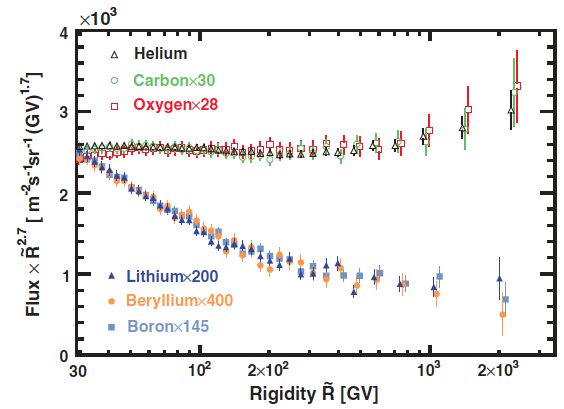}}
        \caption{Comparison of the secondary cosmic ray fluxes~\cite{libeb}
with the AMS primary cosmic ray fluxes~\cite{heco} multiplied by $R^{2.7}$
with their total error as a function of rigidity above 30 GV. For
display purposes only, the C, O, Li, Be, and B fluxes were
rescaled as indicated. For clarity, the He, O, Li, and B data points
above 400 GV are displaced horizontally. } 
\label{exp:fig3}
    \end{center}
\end{figure}
Figure~\ref{exp:fig3} shows the primary and secondary CRs fluxes as a function of rigidity. 
The magnitude and the rigidity dependence of the Li,
Be, and B spectral indices are nearly identical, but distinctly
different from the rigidity dependence of the He, C, and O
spectral indices. 
It is clear that the flux of CRs in the GeV to TeV range cannot be described by a single power law, moreover the
 rigidity behavior between primary and secondary species is remarkably different.
  In particular, above 200 GV, the secondary CRs
harden more than the primaries, pointing to the non universality of spectral indices. 

\begin{figure}[htb]
    \begin{center}
        {\includegraphics[scale=0.5]{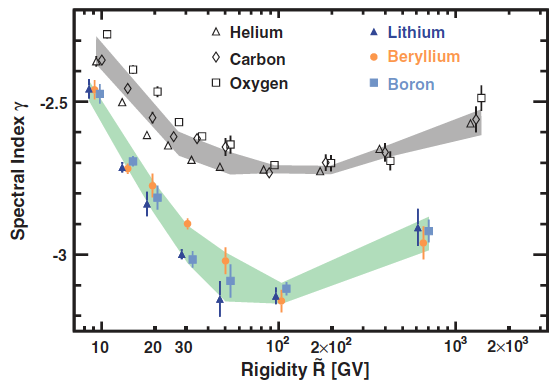}}
        \caption{The dependence of the Li, Be, and B spectral indices on
rigidity, together with the rigidity dependence of the He, C, and O
spectral indices~\cite{libeb}.}
\label{exp:fig4}
    \end{center}
\end{figure}

Figure~\ref{exp:fig4} shows the spectral index $\gamma$ for primary (He, C, O) and secondary CR particles (Li,Be,B) as a function of rigidity,
between 5 GV and 2 TV. For clarity, the Li, B, He, and O data points
are displaced horizontally. 
 
\section{Possible scenarios behind the spectral features}
The uncertainties on current CR measurements up to the knee are significantly smaller than those of measurements carried out one decade ago. 
The new observations revealed 
subtle and unexpected spectral features that require to re-examine the theoretical framework used
to describe the CR origin and propagation. In order to reproduce the observed transition in the spectral index above 200 GeV, we need the transition to arise either at the source (injection or acceleration) 
either during the propagation in the Milky Way. A short description of the open scenarios will be given in the following.  
\subsection{Propagation effects}
One class of scenarios connects the observed spectral features with the description of the propagation in the Milky Way.
In the conventional ``diffusion-convection-reacceleration'' scenario~\cite{galrev} the diffusion coefficient is described
as a single power law in rigidity, and it is space-independent. 
A more complex rigidity dependence of the diffusion coefficient is naturally provided by the non-linear
effects of propagation presented in ~\cite{blasi2} and recently rivisited in~\cite{Evoli}. 
Different explainations for the spectral feature include the  ``two halo model'' discussed in~\cite{tom1} , in which 
CRs are allowed to experience a different type of diffusion when they propagate in the region close to the Galactic disk. 
In the context of this particular model, recent results of a global 
Bayesian analysis based on a Markov-Chain Monte-Carlo sampling algorithm are presented in~\cite{jfeng1}.
\subsection{Source effect}
The space-time discreteness of the galactic sources of CRs, as well as their
 intrinsic features, such as their age or distance, play an important role in the theoretical description of CR fluxes.  
It was proposed that spectral features may be caused by the accidental proximity of a source~\cite{2012MNRAS.421.1209T}. However, this possibility was 
estimated to be extremely faint in~\cite{Genolini:2016}.
In~\cite{blasi3}~\cite{blasi4} it was argued that these classes of models are often in tension with other complementary
observations, namely their predictions on the anisotropy level of the signal is overestimated~\cite{blasi4} and they predict a small diffusion effect, that is not
in agreement with  other measurements, like
the boron to carbon flux ratio. 


The spectral features in proton and helium fluxes are connected with the rise in the positron fraction 
in the frame of a two-component SNR scenario in~\cite{tom2}. The low-energy component (below about 500 GeV) in 
the proton and helium spectrum would be thus a local phenomenon. Nuclei and antiprotons
 would not show a corresponding rise since dominated by a Galactic ensemble
 of SNRs, that are on average younger and
 more efficient to accelerate primary hadrons at high energies (but unable to accelerate secondaries).
 
\subsection{Re-acceleration effects} 
The re-acceleration of both primary and secondary CRs in shock fronts, discussed 
in~\cite{blasi5} and recently revised in~\cite{blasi6} is another viable option to explain the spectral features in galactic CRs.
This model is based on the assumption that the same shocks that are accelerating CR in the sources are also able to re-accelerate the
secondary particles and nuclei that happen to be in the vicinity of the acceleration regions where SN explosions take place. 


\subsection{Secondary to primary ratios as a tool to discriminate between the available scenarios}

The most promising tool to investigate the origin of the spectral features is the study
of secondary species, like boron or lithium, or alternatively on the secondary-to-primary flux ratios, such as boron to carbon (B/C) and complementary observables, like Li/O.
If the transition in the spectral index is already present in the spectra accelerated
at the sources, the B/C should appear featureless, since the spectral feature is conveyed from the parent to the daughter nucleus. 
Conversely, if these features are due to propagation
phenomena, a spectral feature should appear visible in a secondary over primary ratio, being roughly twice as pronounced in secondary species.
Using cosmic-ray boron to carbon ratio (B/C) data recently released by the
AMS-02 experiment~\cite{bc2016}, the first evidence in favor of a diffusive propagation origin
for the broken power-law spectra of protons and helium nuclei was found~\cite{crac4}, by comparing a model with a single power law diffusion coefficient $K(R)$
with a model with a break in $K(R)$. The same number of parameters were used in both cases, deriving the break parameters from the proton and helium fluxes from 
AMS-02 data. A recent independent analysis~\cite{Reinert:2017aga} corroborated this result.

Updated results from AMS-02~\cite{libeb} also including additional 
flux measurement of other secondary species could consolidate this statement. 

\section{Conclusions}
The new observations from recent space borne CR experiments, like PAMELA and AMS-02, revealed 
subtle and unexpected spectral features that require to re-examine or at least improve the theoretical framework used
to describe the CR origin and propagation. In these proceedings I gave a short overview of the recent results of CR proton, helium 
and heavier nuclei up to oxygen: these measurements have a common feature, namely the presence of 
unexpected and still not yet fully understood spectral features.
The main classes of plausible scenarios to interpret the presented results were outlined: the secondary to primary flux ratio, like the B/C, constitute 
a solid observable to ascertain the origin of galactic CRs. If the contribution from local sources seems disfavoured by the anisotropies and by the 
measurement of B/C and other secondary-to-primary flux ratios, the competition
between the propagation effect and the re-acceleration in the vicinity of the shockwaves is not yet concluded. 
A conclusive model that coherently describes the numerous and very precise measurements provided by AMS-02 in the past 7 years is eagerly awaited by the scientific
community. 
\section{Acknowledgements}
I am grateful to the organizers of the XVII Vulcano Workshop for their kind invitation to talk about such a stimulating topic. 
I would like to thank Mathieu Boudaud for his constructive comments and suggestions in the final stage of the preparation of this manuscript. 

%

\begin{thebibliography}{99}
\bibitem{NS} B.~P.~Abbott {\it et al.}, Phys.\ Rev.\ Lett.\  {\bf 119}, 161101 (2017).
\bibitem{NSmm} B.~P.~Abbott {\it et al.}, Astrophys.\ J.\  {\bf 848} (2017) no.2,  L12
\bibitem{IC} M.G. Aartsen {\it et al.}, Science 361, 147-151 (2018).
\bibitem{nucr} M.~L.~Ahnen {\it et al.},to appear in  Astrophys.\ J.\ Lett.\, arXiv:1807.04300 
\bibitem{rev} 
  R.~Aloisio, P.~Blasi, I.~De Mitri and S.~Petrera,
  arXiv:1707.06147.
  \bibitem{blasiRev} 
    P.~Blasi,
    Astron.\ Astrophys.\ Rev.\  {\bf 21}, 70 (2013)

	\bibitem{Serpico2015} 
	  P.~D.~Serpico,
	  PoS ICRC {\bf 2015}, 009 (2016)
	  \bibitem{Drury2017} 
	    L.~O'C.Drury,
	    PoS ICRC {\bf 2017}, 1081 (2018)
	  \texttt{arXiv:1509.04233} 
		\bibitem{PAMELApHe} 
		  O.~Adriani {\it et al.},
		  Science, vol. 332, pp. 69-72, (2011).

		\bibitem{cream} H. S. Ahn {\it et al.},
		Astrophys.\ J.\, {\bf 714}, L89-L93 (2010).

		\bibitem{proton} 
		  M.~Aguilar {\it et al.},
		  Phys.\ Rev.\ Lett.\  {\bf 114}, 171103 (2015).
		\bibitem{heliumF} 
		  M.~Aguilar {\it et al.},
		  Phys.\ Rev.\ Lett.\  {\bf 115}, no. 21, 211101 (2015).

		
 	\bibitem{heco} 
		    M.~Aguilar {\it et al.},
		    Phys.\ Rev.\ Lett.\  {\bf 119}, no. 25, 251101 (2017).
 \bibitem{libeb} 
		  M.~Aguilar {\it et al.},
		  Phys.\ Rev.\ Lett.\  {\bf 120}, no. 2, 021101 (2018).
\bibitem{galrev} 
  A.~W.~Strong, I.~V.~Moskalenko and V.~S.~Ptuskin,
  Ann.\ Rev.\ Nucl.\ Part.\ Sci.\  {\bf 57}, 285 (2007)	
\bibitem{blasi2} P.~Blasi and E.~Amato and P.~D.~Serpico, Phys.\ Rev.\ Lett.\  {\bf 109}, 061101 (2012)
\bibitem{Evoli} C.~Evoli, P.~Blasi, G.~Morlino and R.~Aloisio,  Phys.\ Rev.\ Lett.\  {\bf 121}, no. 2, 021102 (2018)

\bibitem{tom1} N.~Tomassetti, Astrophys. J. {\bf 752}, L13 (2012) 
\bibitem{jfeng1} 
  J.~Feng, N.~Tomassetti and A.~Oliva,
  Phys.\ Rev.\ D {\bf 94}, no. 12, 123007 (2016)
\bibitem{2012MNRAS.421.1209T} Thoudam, S. and H{\"o}randel, J.~R.\ 2012, Mon.\ Not.\ Roy.\ Astron.\ Soc.\, 421, 1209 
\bibitem{Genolini:2016} Y.~Genolini, P.~Salati, P.~Serpico and R.~Taillet, Astron.\ Astrophys.\  {\bf 600}, A68 (2017)

\bibitem{blasi3} P.~Blasi and E.~Amato, \ 2012, JCAP, 1, 010 
\bibitem{blasi4} P.~Blasi and E.~Amato, E.\ 2012, JCAP, 1, 011

\bibitem{tom2} N.~Tomassetti and F.~Donato,	Astrophys.\ J.\  {\bf 803}, no. 2, L15 (2015)
\bibitem{blasi5} P.~Blasi, G.~Morlino, R.~Bandiera, E.~Amato, \& D.~Caprioli, Astrophys.\ J.\, {\bf 755}, 121 (2012)
\bibitem{blasi6} P.~Blasi,  Mon.\ Not.\ Roy.\ Astron.\ Soc.\  {\bf 471}, no. 2, 1662 (2017)
\bibitem{bc2014} 
  O.~Adriani {\it et al.},
  Astrophys.\ J.\  {\bf 791}, no. 2, 93 (2014)
\bibitem{bc2016} 
  M.~Aguilar {\it et al.},
  Phys.\ Rev.\ Lett.\  {\bf 117}, no. 23, 231102 (2016).
\bibitem{crac4} 
  Y.~Genolini {\it et al.},
  Phys.\ Rev.\ Lett.\  {\bf 119}, no. 24, 241101 (2017)
\bibitem{Reinert:2017aga} 
  A.~Reinert and M.~W.~Winkler,
  JCAP {\bf 1801}, no. 01, 055 (2018)
\end{thebibliography}
\end{document}